\documentstyle[preprint,aps]{revtex}
\advance\hsize by 0.0truecm  \hoffset=0.7cm
\begin{document}
 \ifx\fiverm\undefined
         \newfont\fiverm{cmr5}
  \fi
\input pictex
\renewcommand{\baselinestretch}{0.67}
\bibliographystyle{prsty}
\newcommand{\dirac}{\!\!\!\not{\!\partial}}
\newcommand{\srG}{2\sqrt{G(G+1)}}
\newcommand{\C}{\cite}
\newcommand{\beq}{\begin{equation}}
\newcommand{\eeq}{\end{equation}}
\newcommand{\bea}{\begin{eqnarray}}
\newcommand{\eea}{\end{eqnarray}}
\newcommand{\bit}{\begin{itemize}}
\newcommand{\eit}{\end{itemize}}
\newcommand{\ms}{m_{\rm s}}
\newcommand{\MeV}{{\rm MeV}}
\newcommand{\fm}{{\rm ~fm}}
\newcommand{\alf}{{\bar a}}
\newcommand{\bet}{{\bar b}}
\newcommand{\lb}{\hfil\break }
\newcommand{\qeq}[1]{eq.\ (\ref{#1})}
\newcommand{\dsl}{ \rlap{/}{\partial} }
\newcommand{\pst}{ \rlap{/}{p}  }
\newcommand{\half}{\frac{1}{2}}
\newcommand{\quref}[1]{\cite{bolo:#1}}
\newcommand{\qref}[1]{Ref.\ \cite{bolo:#1}}
\newcommand{\queq}[1]{(\ref{#1})}
\newcommand{\qtab}[1]{Tab. \ref{#1}}
\newcommand{\wu}{\sqrt{3}}
\newcommand{\nn}{\nonumber \\ }
\newcommand{\Y}{\ {\cal Y}}
\newcommand{\Sp}{{\rm Sp\ } }
\newcommand{\Spto}{{\rm Sp_{(to)}\ } }
\newcommand{\Tr}{{\rm Tr\ } }
\newcommand{\tr}{{\rm tr\ } }
\newcommand{\sign}{{\rm sign} }
\newcommand{\linie}{\ \vrule height 14pt depth 7pt \ }
\newcommand{\intT}{\int_{-T/2}^{T/2} }
\newcommand{\Nc}{N_{\rm c}}
\newcommand{\Ne}{$N_{\rm c}$}
\newcommand{\gs}{$g_{\rm A}^{(0)}$}
\newcommand{\gt}{$g_{\rm A}^{(3)}$}
\newcommand{\go}{$g_{\rm A}^{(8)}$}
\newcommand{\Gs}{g_{\rm A}^{(0)}}
\newcommand{\Gt}{g_{\rm A}^{(3)}}
\newcommand{\Go}{g_{\rm A}^{(8)}}
\newfont{\exn}{cmssi10 scaled 1200}
\newcommand{\vx}{{\vec x}}
\newcommand{\vy}{{\vec y}}
\newcommand{\vz}{{\vec z}}
\newcommand{\val}{{\rm val}}
\newcommand{\ba}{\begin{array} }
\newcommand{\ea}{\end{array} }
\newcommand{\disp}{\displaystyle}
\preprint{   $  \begin{array}{r}
  \disp \rm  SUNY-NTG-95-13  \end{array}    $   }

\title{The Constituent Quark Limit and the Skyrmion Limit \\  of
Chiral Quark Soliton Model}
\renewcommand{\baselinestretch}{0.9}
\author{
Micha{\l} Prasza{\l}owicz$^{(1)}$
\footnote{supported by Alexander von Humboldt Foundation \\
\phantom{1} email:~michal@thrisc.if.uj.edu.pl},
Andree Blotz$^{(2,3)}$
 \footnote{supported by Alexander von Humboldt Foundation \\
\phantom{1}   email:~andreeb@luigi.physics.sunysb.edu}
and Klaus Goeke$^{(3)}$
\footnote{email:~goeke@hadron.tp2.ruhr-uni-bochum.de}  }
\bigskip
\bigskip
\address{(1)  Institute of Physics,
Jagellonian University, Reymonta 4, \\30-059~Krak{\'o}w, Poland}
\bigskip
\bigskip
\address{(2) Department of Physics, State University of New York at Stony
Brook, \\
Stony Brook, New York, 11794, USA \\ }
\bigskip
\address{(3)
Institute for  Theoretical  Physics  II, \\  P.O. Box 102148,
Ruhr-University Bochum, \\
 D-W-44780 Bochum, Germany  \\       }
\bigskip
\date{\today}
\renewcommand{\baselinestretch}{0.9}
\maketitle

\begin{abstract}
We calculate $\Gt$ and $\Gs$ in the Chiral Quark-Soliton Model (or
equivalently in semibosonized Nambu--Jona-Lasinio model) in the limit
of small and large soliton size. In small soliton limit we recover the
results of the Constituent Quark Model. The agreement between the two
models is achieved due to the recently calculated $1/\Nc$ contributions
to $\Gt$ and to $\Gs$. In the case of $\Gt$ these terms arise from the
{\it time-ordering} of the collective operators. For large solitons
simple scaling arguments show that the $1/\Nc$ contributions vanish
with the inverse power of the soliton size.
\end{abstract}
\renewcommand{\baselinestretch}{1}
\vfill\eject
\renewcommand{\arraystretch}{0.5}

The Constituent Quark Model (CQM), although often considered as naive,
has been often used to calculate {\em a priori} dynamical hadronic
matrix elements  in terms of simple group-theoretical factors. The most
prominent prediction of the  CQM, namely the value of the singlet axial
constant \gs=1, caused the so-called "spin crisis". The CQM prediction
\quref{karlpat}
for $\Gt=(\Nc+2)/3$, which overshoots the experimental value of 1.26,
triggered the interest in the possible renormalization of a quark axial
coupling away from 1
\C{bolo:ManGeor,bolo:Wein1}.
On the other extreme Skyrme-like models, in which
baryons are considered as nontrivial configurations of the classical
pion-sigma chiral fields, gave numbers in flat contradiction with the
CQM (and experiment): $\Gt<1$ \quref{anw}
and $\Gs=0$ \quref{bek1}.  These models are subjected
to a semiclassical quantization of the rotational zero-modes -- only
then the baryon spectrum is reproduced \quref{anw}.  The class of {\em
interpolating} models, like the quark Nambu--Jona-Lasinio (NJL) model
(or equivalently Chiral Quark-Soliton Model)
\C{bolo:njl1,bolo:dipepo,bolo:rewu,bolo:DiPePr,bolo:megrgo,bolo:wayo},
in which both quarks and chiral fields are present, seemingly did not
improve upon the $g_{\rm A}$'s, being on the low side in rough
agreement with the Skyrme model. However not only the values of the
$g_{\rm A}$'s have been wrong, but also the \Ne ~counting. In the NJL
model and in the Skyrme model as well, $\Gt$ scales like \Ne,
{}~however the ${\cal O}(1)$ terms, which in the CQM are of utmost
importance, seemed to vanish identically. The same concerns the ${\cal
O}(1)$ contributions to $\Gs$ which in the CQM sacles like 1.

Recently ${\cal O}(1)$ corrections to the $g_{\rm A}$'s and to the
magnetic moments have been calculated in the NJL model
\C{bolo:wawa,bolo:ab9,bolo:chr2,bolo:ab10}
and also in the chiral bag model \C{bolo:hoto}.
Not only have they significantly improved the numerics
of the NJL model, but also the \Ne ~scaling, as predicted by the CQM,
has been recovered. On the contrary in the Skyrme-like models based on
a local effective meson lagrangian these corrections vanish identically
Ref.\C{bolo:schewe}.

The origin of these -- for a long time overlooked -- corrections
(as has been shown in detail in Refs.\C{bolo:ab9,bolo:chr2,bolo:ab10})
is twofold : 1) the NJL model is formulated in terms of a {\em
non-local} effective action, namely in terms of the fermion determinant,
and the operators appearing in the path integral for matrix elements
have to be {\it time-ordered}, 2) the seemingly commuting c-number
quantities like angular velocities and the Euler angles in fact do not
commute, since upon the semiclassical quantization they are promoted to
quantum operators. Apart from these {\it time-ordered} corrections
there are also terms which come from  the {\it anomalous} (or
imaginary) part of the effective action (as in the case of $\Gs$). In
this order of the $1/\Nc$ expansion their emergence does not depend on
the time order of the collective operators.

In this short note we show, that in the small soliton limit of the NJL
model the results of the CQM for the $g_{\rm A}$'s are recovered
indepently on the model parameters. Indeed, when the soliton ceases to
exist the quarks, which are normally bound in a potential created by
the chiral fields, reach the plane wave orbits of the free Dirac
particles \C{bolo:DiPePr}. Moreover, the contribution of the Dirac sea
(or equivalently of the chiral fields) vanishes in this limit.
Strictly speaking the Constituent Quark Limit can be accomplished by
tuning the constituent quark mass $m\rightarrow 0 $, in such a way that
for each $m$ the selfconsistent soliton solution is obtained. Since for
some critical $m_{\rm cr}$ the bound state solution ceases to exist even if
the soliton has finite size, we choose first to push $r_0 \rightarrow 0 $
{\it artificially} for some unspecified $m>m_{\rm cr}$, for which the
selfconsistent soliton solution does exist. It is then enough to observe,
that the result does not depend on $m$.


Furthermore, on the basis of a simple scaling argument we show,
that in the large soliton limit the new corrections die out. Therefore
in this limit the predictions of local mesonic effective theories are
parametrically reproduced. The picture which emerges is therefore very
appealing: the small and large soliton limits of the NJL model
coincide
with the previously obtained results within the CQM and the Skyrme model.
Were the new {\it non-local} ${\cal O}(1)$ corrections equal to zero,
the nice correspondence with the CQM would be completely lost.

Let us first briefly remind the philosophy and main ingredients of the
NJL model.  Our  starting point is a semibosonized Euclidean action of
the NJL model \C{bolo:ab9,bolo:ab10}:
\beq
S_{\rm eff}[U]=-{\rm Sp} \log
\left\{ -i~~\dirac \, +\, m \; U^{\gamma_{5}} \right\},
\label{eq:Seff}
\eeq
where $\gamma$ matrices are antihermitian, the SU(2) matrix
$U^{\gamma_{5}}=\exp(i \gamma_5\vec\tau \vec\pi/F_{\pi})$
describes chiral fields and $m$ is the constituent
quark mass.  Eq.(\ref{eq:Seff}) has to be  regularized.  For the
hedgehog Ansatz $ {U}_{0}^{\gamma_5}= \cos P(r) + i \vec{n}  \vec{\tau}
\gamma_5 \sin P(r)
$ with the profile function satisfying the boundary conditions
$P(0)=\pi$ and $P(\infty)=0$, the model has solitonic solutions.
The existence of the soliton is achieved due to the interplay between
the valence level and the energy of the continuum states. All
quantities have therefore typically two contributions which we
subsequently call {\it valence} and {\it sea} respectively.

If the soliton size $r_0$ (defined through $P=P(r/r_0)$) decreases, the
valence level joins the upper continuum and the {\it sea} contribution
vanishes. For large solitons the valence level sinks into the Dirac sea
and the explicit contribution of the valence level disappears. In this
way the model interpolates between the CQM and the Skyrme-like models
\footnote{One should stress, however, that the four-derivative term
from the NJL model does not coincide with the one of the Skyrme model.
It is cutoff sensitive and it can stabilize the soliton only for very
large constituent masses \quref{megrgo}.}.

The energy of the soliton is given as the regularized sum of the
eigenenergies of the Dirac hamiltonian:
\beq
H(U_0)=-i \gamma_4 (-i \gamma_k \partial_k + m U_0^{\gamma_5}). \label{Dham}
\eeq
H commutes neither with the isospin ($T$), the spin ($S$), nor with the
angular momentum ($L$), but only with the so called grand-spin
$G=T+J=T+L+S$.

To appreciate the way the constituent quark limit of the NJL
model is actually reached, requires a careful analysis of the
wave functions of the hamiltonian \queq{Dham}. To this end
one defines spin-isospin spherical harmonics \C{bolo:kari}:
\bea
\Xi^{(\pm,\pm)}_{G,M} &=&
\langle \theta,\phi\mid G,M;j=G\pm\half,l=G\pm1 \rangle, \nn
\Xi^{(\pm,\mp)}_{G,M} &=&
\langle \theta,\phi \mid G,M;j=G\pm\half,l=G~~\rangle,
\label{harm} \eea
where $G,~M,~j$ and $l$ denote the eigenvalues of the grand-spin and
its third component and of the total momentum and the angular momentum
respectively.  The soliton Dirac equation splits naturally into two
parity ${\cal P}=\Pi (-)^{G}$ sectors.  Natural parity $\Pi=(+)$ spinors
are given by:
\beq
 \Psi^+_{G,M}(x)=
 \left(  \ba{c}  i A_G^+(r) \;\Xi^{(+,-)}_{G,M}
            +  i C_G^+(r) \;\Xi^{(-,+)}_{G,M}    \\
                          B_G^+(r) \;\Xi^{(+,+)}_{G,M}
           +    D_G^+(r) \;\Xi^{(-,-)}_{G,M}
                 \ea   \right),
        \label{spip}
\eeq
whereas for $\Pi=(-)$ they take the following form:
\beq
  \Psi^-_{G,M}(x)=
  \left(  \ba{c} - i B_G^-(r) \;\Xi^{(+,+)}_{G,M}
            -  i D_G^-(r) \;\Xi^{(-,-)}_{G,M}    \\
                      ~~A_G^-(r)  \;\Xi^{(+,-)}_{G,M}
           +    C_G^-(r) \;\Xi^{(-,+)}_{G,M}
                 \ea   \right).
          \label{spin}
\eeq
Then, for $\Pi=(+)$, the Dirac equation is reduced to a system of 4
first order differential equations for the radial functions $A\ldots
D$:
\bea
\frac{d}{dr} \left[ \ba{c} A \\ B \\ C \\ D \ea \right] & = &
\left[ \ba{cccc} G/r & m\cos P+E & 0 & 0 \\
              m\cos P-E& -(G+2)/r & 0 & 0 \\
               0 & 0 &-(G+1)/r& m\cos P+E  \\
               0 & 0 & m\cos P-E& (G-1)/r \ea \right]
               \left[ \ba{c} A \\ B \\ C \\ D \ea \right] \nn
 & & \nn
& + & \frac{m \sin P}{2G+1}
\left[ \ba{cccc} -1 & 0 &~~b_G~~ & 0 \\
                  0 & 1 &  0  &~-~b_G~~ \\
                 ~~b_G~~ & 0 & 1 & 0 \\
                  0 &~-~b_G~~ & 0 & -1   \ea \right]
               \left[ \ba{c} A \\ B \\ C \\ D \ea \right], \label{equ}
\eea
where $b_G=\srG$. Grand spin subscripts $G,M$ and parity superscript
"+" have been suppressed.  For $\Pi=(-)$ one should change the
sign of $m$ in \queq{equ}.
For any  soliton shape one can choose two different boundary conditions
in the origin: 1) $A\sim r^{K},~~B \sim r^{K+1} $ and $C=D=0$ or
2) $A=B=0$ and $C\sim r^{K},~~D \sim r^{K-1}$. Therefore the solutions are
labelled by a multiindex $n=(n,G,M,\Pi,i)$, where $n$ stands also for a
radial quantum number and $i$ corresponds to the initial conditions
discussed above.

In order to quantize the model one performs collective quantization in
terms of the rotation matrix $A(t)$: $U^{\gamma_5}(x) = A (t)
U_{0}^{\gamma_5} (\vec{x}) \,
A^{\dagger}(t) $, where $ A^{\dagger}\; dA/dt = i/2\;  \sum_{a=1}^{3}
\tau_{a} \Omega_{a}$ \quref{anw}. As a result one gets a {\em
collective} hamiltonian:
\beq
{\cal H}_{\rm coll} = M_{\rm sol} + \frac{{\vec{J}}^2}{2I}. \label{Hcoll}
\eeq
Here $M_{\rm sol}$ is the soliton mass and $I$ the soliton moment of
inertia. Spin operators $\vec J$ are related to the angular velocities
by the quantization prescription: $I \vec{\Omega}\rightarrow \vec{J}$.
The eigenstates of \queq{Hcoll} have been successfully interpreted as
baryons. Their wave functions are given in terms of $D^{(1/2)}(A)$
Wigner matrices  and the moment of inertia $I$ is given as the sum
over the eigenstates of the {\it intrinsic} Dirac hamiltonian
\queq{Dham} (in contrast to the {\it collective} hamiltonian
\queq{Hcoll}). Only the {\it valence} part of $I$ is here of interest:
\beq  I_\val =  {N_c\over 6}
               \sum_{n\ne\val}
                 {  \langle \val \mid  \vec{\tau}
                  \mid n \rangle  \langle n \mid
                     \vec{\tau}   \mid \val  \rangle \over
                          E_n - E_\val   }.
\label{i1}
\eeq
Let us note in passing that since $I\sim\Nc$ each power of $\Omega$
counts as one power of $1/\Nc$.

Using the same path-integral formalism, which has been employed to
derive ${\cal H}_{\rm coll}$ one can derive a compact expression for
any quark operator of interest. The space-like collective axial current
operator can be expressed by the following functional  trace in
Euclidean space \quref{ab9}:
\beq
A_{j}^{a}=i\; {\rm tr} \left\{ \Gamma_{j}^{b} \; D_{a b}
            \frac{1}{\partial_t+H(U_0^{\gamma_5})+\frac{i}{2} \Omega_c^{\rm E}
            \lambda_c} \right\}
\label{eq:Jj} \eeq
with $\Gamma_{j}^{b}=\gamma_4 \gamma_j \gamma_5 \tau_b$ and
$D_{ab}=1/2\;{\rm Tr}(A^\dagger \tau_a A \tau_b)$. For a singlet axial
current $ \tau_b$ should be replaced by a unit matrix and one should
also replace $D_{a b} \rightarrow 1_{ab}$ in eq.(\ref{eq:Jj}).
It is implicitly understood that the {\it sea} part of eq.(\ref{eq:Jj})
is regularized and that the vacuum contribution, corresponding to
$U_0^{\gamma_5} \rightarrow 1$ is subtracted.

Let us expand eq.(\ref{eq:Jj}) in $\Omega$, remembering that each power
of $\Omega$ counts as $1/\Nc$. In the zeroth order we get:
\beq
(A_{j}^{a})^{(0)}=
i\; {\rm tr} \left\{ \Gamma_{j}^{b} \; D_{a b}
            \frac{1}{\partial_t+H} \right\}
\equiv  D_{aj}\; \alpha.
\label{eq:Jj0}\eeq
Note that at this level the singlet axial current is equal to 0 in
agreement with the Skyrme model result \C{bolo:bek1}.

The next term in $\Omega$ (or in $1/\Nc$ ) is  of great importance.
Expanding in $\Omega$ we get:
\beq
(A_{j}^{a})^{(1)}=\frac{1}{2}\; {\rm tr} \left\{ \Gamma_{j}^{b} \; D_{a b}
            \frac{1}{\partial_t+H} \Omega^{\rm E}_c
            \lambda_c \frac{1}{\partial_t+H}
       \right\}. \label{eq:Jj1}
\eeq
There are two subtleties connected with eq.(\ref{eq:Jj1}): first it
should be remembered that because of the collective quantization
$\Omega_c$ is no longer a c-number but rather an operator $J_c/I$ which
does not commute with $D_{a b}$.  Second, the trace in
eq.(\ref{eq:Jj1}) should be understood as time-ordered
\C{bolo:ab9,bolo:chr2,bolo:ab10}.

To this end let us consider the Euclidean propagator:
\bea
<x \mid \frac{1}{\partial_t + H} \mid y > & = &
\theta(t_x-t_y)\sum\limits_{E_n>0} \Phi_n(\vec{x})\Phi_n^{\dagger}(\vec{y})
\exp(-E_n(t_x-t_y)) \nonumber \\
 & - &\theta(t_y-t_x)\sum\limits_{E_n<0}
 \Phi_n(\vec{x})\Phi_n^{\dagger}(\vec{y}) \exp(-E_n(t_x-t_y)). \label{eq:prop}
 \eea
If two such propagators are multiplied and time ordering in
eq.(\ref{eq:Jj1}) is assumed, then the correct expression for the baryon number
one solutions  reads
(back in the Minkowskian space):\bea
(A_{j}^{a})^{(1)} & = &  \frac{i}{2 I}\;
\sum\limits_{E_k<\mu \atop E_n > \mu}
\frac{1}{E_k -E_n}
\left\{
<k \mid \Gamma_{j}^{b} \mid n > < n \mid \tau_c \mid k >
 D_{a b} J_c \right.
  \nonumber \\
 & &
 \hspace{3 cm} + \left.
 <n \mid \Gamma_{j}^{b} \mid k > < k \mid \tau_c \mid n >
J_c D_{a b}
\right\}, \label{eq:OK}
\eea
where the chemical potential $E_\val<\mu<m$.
Note that the order of $D_{ab}$ and $J_c$ is unambiguously dictated
by the time ordering of the path-integral. One can always choose the
basis $|\tilde{n}>$ (by a transformation of $|n>$'s in each $G$
sector separately) such, that:\bea
 <\tilde{n} \mid \Gamma_{j}^{b} \mid \tilde{k} >
 =  <\tilde{k} \mid \Gamma_{j}^{b} \mid \tilde{n} >, &~~~~~ &
  <\tilde{n} \mid \tau_c \mid \tilde{k} >
= -   <\tilde{k} \mid \tau_c \mid \tilde{n} >. \label{sym}
\eea
If at this point one does not pay attention to the order of $D_{ab}$
and $J_c$ in eq.\queq{eq:OK}, one can easily get a zero answer for
$(A_{j}^{a})^{(1)}$ in view of the symmetry properties of
eq.\queq{sym}. However, keeping the right order of the collective
operators and making use of the property $[J_a,D_{jb}]=i\epsilon_{abc}
D_{jc}$, one finally arrives at the following expressions for the
collective axial constants operators:
\bea
g_{\rm A}^{(3)} =  -\left(\alpha+\frac{\beta}{I}\right)\; D_{33},
&~~~~~~~~~& g_{\rm A}^{(0)} =~\frac{\gamma}{I}\; J_3.
\label{ggA}
\eea
Note that the $1/\Nc$ correction corresponding to $\beta/I$ has the
same group-theoretical factor $D_{33}$ as the leading term in
accordance with the general theorem of Ref.\C{bolo:dama2}
The {\it valence} part of $\alpha$ entering eq.\queq{ggA} and
following from eq.\queq{eq:Jj0} is given
by:
\beq
  \alpha_\val = {N_c\over 3}
     \langle \val \mid {\vec\sigma}{\vec\tau} \mid \val
          \rangle.
      \label{ax1}
\eeq
Furthermore from eq.\queq{eq:OK}:
\bea  \beta_\val & = &
   -  {N_c\over 6}\; i  \epsilon_{ijk}
\sum_{n\ne\val}
        { {\langle \val \mid  \sigma_i \tau_j
         \mid n \rangle  \langle n \mid
             \tau_k   \mid  \val \rangle} \over
                { E_n - E_\val}   } \ \sign E_n ,
 \\          \label{ax2}
\gamma_\val &=&   -  {N_c\over 6}
           \sum_{n\ne\val}
        { \langle \val \mid  \vec{\sigma}
         \mid n \rangle  \langle n \mid
             \vec{\tau}   \mid  \val \rangle \over
                 E_n - E_\val   }.
        \label{ax3}
\eea

Now we will proceed to define the constituent quark limit more
precisely. For the sake of simplicity we will use a somewhat unphysical
profile function $P=\pi \theta(r_0-r)$ and then tune $r_0 \rightarrow
0$. In this limit the energy of the valence level $E_\val \rightarrow
m$ and the {\it sea} contributions to all quantities vanish.
Simultaneously we will take the nonrelativistic limit, i.e. we will
systematically neglect small components of the spinors \queq{spip} and
\queq{spin}. As we shall illustrate later the results are not changed
for a realistic smooth profile function.

For the $\theta$  profile the term proportional to $\sin P$ in
eq.\queq{equ} vanishes and the radial functions $C(r)\equiv D(r)\equiv
0$ for {\it all} $r$, for boundary conditions $i=1$. Similarly
$A(r)\equiv B(r)\equiv 0$ for $i=2$.  Therefore for given energy $E_n$,
$G$ and $M$ we have four independent solutions corresponding to two
parities and two boundary conditions:
\bea
\langle x \mid n,G,M,+,1 \rangle =
\left( \ba{r} ~i~ A^+_{nG} \; \Xi^{(+,-)}_{G,M} \\
                B^+_{nG} \; \Xi^{(+,+)}_{G,M} \ea \right), &~&
\langle x \mid n,G,M,+,2 \rangle =
\left( \ba{r} ~i~ C^+_{nG} \; \Xi^{(-,+)}_{G,M} \\
                D^+_{nG} \; \Xi^{(-,-)}_{G,M} \ea \right),    \nn
\langle x \mid n,G,M,-,1 \rangle =
\left( \ba{c}-i B^-_{nG} \; \Xi^{(+,+)}_{G,M} \\
                A^-_{nG} \; \Xi^{(+,-)}_{G,M} \ea \right), &~&
\langle x \mid n,G,M,-,2 \rangle =
\left( \ba{c}-i D^-_{nG} \; \Xi^{(-,-)}_{G,M} \\
                C^-_{nG} \; \Xi^{(-,+)}_{G,M} \ea \right). \label{4sol}
\eea
For the valence level "$\val$" corresponds to $(n=0,G=0,M=0,\Pi=+,i=1)$.
Note that
$\Xi_{0,0}^{(-,\pm)}$ identically vanish, so for $G=0$ there is no
solution corresponding to the second boundary conditions $i=2$.

The nonrelativistic limit of eq.\queq{equ} can be obtained by
approximating $E\approx m$. Then, for $r > r_0$ $A^+$ and
$C^+$ become large components of natural parity spinors, whereas for
unnatural parity large components are given by $B^-$ and $D^-$. For
$r < r_0$ the upper and lower components are interchanged, since
the sign of the mass term in eq.\queq{equ} flips, once $r$ crosses
$r_0$. Therefore,  the nonrelativistic limit corresponds
to keeping  $B^+=D^+\approx0$ for $r>r_0$ and $A^-=C^-\approx
0$ for $r<r_0$. The normalized solutions can be obtained if the system
is confined in a box of radius $R$.  Then the normalization condition
for the valence level in the nonrelativistic limit reads:
\beq
\int\limits_{0}^{r_0} dr\ r^2 \mid B_{00}^+ \mid^2
+ \int\limits_{r_0}^{R} dr\ r^2 \mid A_{00}^+ \mid^2=1. \label{norm}
\eeq
Obviously, in the limit $r_0 \rightarrow 0$ the first integral has to
vanish and the normalization is given entirely by the second term in
eq.\queq{norm}. One can similarly normalize other solutions
corresponding to the nonvalence orbits.


Now, in order to calculate constants $\alpha_\val$, $\beta_\val$,
$\gamma_\val$ and the valence part of the moment of inertia we have to
know the matrix elements of intrinsic operators between the states of
eq.\queq{4sol}. Constant $\alpha_\val$ can be easily evaluated by means
of the Wigner-Eckart theorem:
\beq
\alpha_\val=- \frac{\Nc}{3} \left(- \int\limits_{0}^{r_0} dr\ r^2 \mid
B_{00}^+ \mid^2 + 3 \int\limits_{r_0}^{R} dr\ r^2 \mid A_{00}^+
\mid^2\right). \label{al}
\eeq
In the limit $r_0 \rightarrow 0$ the first integral vanishes and
$\alpha_\val= -\Nc$, due to the normalization condition \queq{norm}.
The proton-spin-up matrix element of the collective operator $D_{33}$
is equal to $-1/3$, so the leading term of $g_{\rm A}^{(3)}$ is equal
to $\Nc/3$.

In order to evaluate the next term in eq.\queq{ggA} let us first
observe that the matrix elements
$\langle n~|\vec{\tau}|\val\rangle$
do not vanish only for unnatural parity G=1 state
$n=(n,1,M^{\prime},-,2)$ -- this can be again easily verified by the
Wigner-Eckart theorem.  Here $M^{\prime}=0,\pm 1$ depending on whether
we take $\tau_3$ or $\tau_{\pm}$ respectively. Since the ratio
$\beta_\val/I_\val$  is dominated by the lowest energy level which has
the energy $E_0\approx m$ the sums in \queq{ax2} and \queq{i1} reduce
actually only to the one term.  After calculating
$i \epsilon_{ijk} \langle \val | \sigma_i \tau_j |~n \rangle$,
where again $n=(0,1,M^{\prime},-,2)$,
one gets finally:
\bea
\beta_\val & = & \frac{\Nc}{3}
\lim\limits_{E_0\rightarrow m}\frac{1}{E_0-E_\val}
\left( \int\limits_{0}^{r_0} dr\ r^2 B_{00}^+ C_{01}^-
   -   \int\limits_{r_0}^{R} dr\ r^2 A_{00}^+ D_{01}^- \right)  \nn
   & & \hspace{3cm} \times
\left(  \int\limits_{0}^{r_0} dr\ r^2 B_{00}^+ C_{01}^-
   + 3  \int\limits_{r_0}^{R} dr\ r^2 A_{00}^+ D_{01}^- \right).
\label{be} \eea
Similarly:
\beq
I_\val  = \frac{\Nc}{2}
         \lim\limits_{E_0\rightarrow m} \frac{1}{E_0-E_\val}
\left( \int\limits_{0}^{r_0} dr\ r^2 B_{00}^+ C_{01}^-
   -   \int\limits_{r_0}^{R} dr\ r^2 A_{00}^+ D_{01}^- \right)^2.
\eeq
The constituent quark limit (CQL) consists in taking $r_0\rightarrow 0$
and $E_\val \rightarrow m$:
\beq \frac{\beta_\val}{I_\val}\stackrel{\rm CQL}{\longrightarrow} -2.
\eeq
Proceeding in a similar way one gets:
\bea
\gamma_\val & = & -\frac{\Nc}{3}
                 \lim\limits_{E_0\rightarrow m} \frac{1}{E_0-E_\val}
\left( \int\limits_{0}^{r_0} dr\ r^2 B_{00}^+ C_{01}^-
   -   \int\limits_{r_0}^{R} dr\ r^2 A_{00}^+ D_{01}^- \right)  \nn
   & & \hspace{3cm} \times
\left( \int\limits_{0}^{r_0} dr\ r^2 B_{00}^+ C_{01}^-
   +  3 \int\limits_{r_0}^{R} dr\ r^2 A_{00}^+ D_{01}^- \right),
\label{ga} \eea
so that:
\beq \frac{\gamma_\val}{I_\val}\stackrel{\rm CQL}{\longrightarrow} 2.
\eeq
Altogether we find:
\bea
g_{\rm A}^{(3)}\stackrel{\rm CQL}{\longrightarrow} \frac{\Nc+2}{3}
&~~~~~{\rm and}~~~~~&
g_{\rm A}^{(0)}\stackrel{\rm CQL}{\longrightarrow}1
\eea
in perfect agreement with the CQM results. One can convince oneself by a
direct numerical calculation, that this result holds for any realistic
soliton profile, whose size $r_0\rightarrow 0$. This is illustrated in
Fig.1, where the \gt~ dependence on $r_0$ is plotted for $P=2\;{\rm
arctan}((r_0/r)^2)$.


Now, in order to investigate the large soliton limit, it is enough to
observe, that in this limit $\alpha\simeq\beta$ \quref{ab10} in the
sense of radial dependence and
both scale like
$r_0^2$
\C{bolo:ab10},
whereas $I$ scales like $r_0^3$ \quref{ab4}.  Therefore for
$r_0\rightarrow\infty$ the new ${\cal O}(1)$ terms are suppressed and
the results of the Skyrme model are parametrically reproduced.  However
for large but finite $r_0$, the correction  exists and differs from the
Skyrme model in contrast to the chiral bag calculations in \qref{hoto}.
In the case of $g_{\rm A}^{(0)}$ the large soliton limit of $\gamma$
goes as inverse powers of $r_0$ \quref{ab6}, so that it is vanishing
and the Skyrme limit $g_{\rm A}^{(0)}=0$ is reproduced.  So we have in
fact a {\it smooth} interpolation between the CQM and Skyrme model.

Summarizing: the axial coupling constants $\Gt$ and $\Gs$ evaluated in
the Nambu--Jona-Lasinio  model  in a small soliton limit agree with the
Constituent Quark Model. Since the NJL model is a nontrivial dynamical
model with valence quarks and polarized Dirac sea, and the CQM is based
on purely {\it kinematical} assumptions concerning the symmetries of
the wave functions, this agreement is by no means obvious. It is only
possible because of the subtle cancellation of the overlap integrals
of {\it a priori} very different operators. It is also
important that the overlaps {\it inside} the soliton vanish with the
soliton size approaching 0.  Moreover, if the ${\cal O}(1)$ corrections
coming from the {\it time-ordering} of the collective operators were
not taken into account,
the agreement for $\Gt$ would never be possible.
For it is hard to believe that other corrections, like meson loops for
example, which require extra regularization and envisage completely
different dynamics, might bring about the similar cancellation
(although this statement requires obviously quantitative analysis).
Therefore we consider the results of this paper as a further
confirmation of the correctness of the procedure presented in
Refs.\C{bolo:ab9,bolo:chr2,bolo:ab10}, where the detailed derivation of
the ${\cal O}(1)$ rotational corrections has been presented.

\acknowledgements
The authors acknowledge the support of Alexander von Humboldt
Foundation (AB, MP) and Department of Energy  grant
DE-FG02-88ER40388 (AB).

\renewcommand{\baselinestretch}{0.8}

\begin{thebibliography}{10}

\bibitem{bolo:karlpat}
G.Karl and J.E.Paton, Phys. Rev. {\bf D30},  238  (1984).

\bibitem{bolo:ManGeor}
A.Manohar and H.Georgi, Nucl.Phys. {\bf B234},  189  (1984).

\bibitem{bolo:Wein1}
S. Peris, Phys. Lett. {\bf B268}, 415 (1991).

S. Weinberg, Phys. Rev. Lett. {\bf 67},  3473  (1991).

\bibitem{bolo:anw}
G. Adkins, C. Nappi, and E. Witten, Nucl. Phys. {\bf B228},  552  (1983).

\bibitem{bolo:bek1}
S. Brodsky, J. Ellis, and M. Karliner, Phys. Lett. {\bf B206},  309  (1988).

\bibitem{bolo:njl1}
Y. Nambu and G. Jona-Lasinio, Phys. Rev. {\bf 122},  345  (1961).

\bibitem{bolo:dipepo}
D.Diakonov, V.Petrov, and P.Pobylitsa, Nucl. Phys. {\bf B306},  809  (1988).

\bibitem{bolo:rewu}
H.Reinhardt and R.Wuensch, Phys. Lett. {\bf B215},  577  (1988).

\bibitem{bolo:DiPePr}
D.Diakonov, V.Petrov, and M.Prasza{\l}owicz, Nucl. Phys. {\bf B323},  53
  (1989).

\bibitem{bolo:megrgo}
Th.Meissner, F.Gruemmer, and K.Goeke, Phys. Lett. {\bf B227},  296  (1989).

\bibitem{bolo:wayo}
M. Wakamatsu and H. Yoshiki, Nucl. Phys. {\bf A524},  561  (1991).

\bibitem{bolo:wawa}
M. Wakamatsu and T. Watabe, Phys. Lett. {\bf B312},  184  (1993).

\bibitem{bolo:ab9}
A.Blotz, M. Prasza{\l}owicz, and K. Goeke, Phys. Lett. {\bf B317},  195
  (1993).

\bibitem{bolo:chr2}
C. Christov {\it et~al.}, Phys. Lett. {\bf B325},  467  (1994).

\bibitem{bolo:ab10}
A.Blotz, M. Prasza{\l}owicz, and K. Goeke, RUB report No. TPII-41/93, 1993.

\bibitem{bolo:hoto}
A. Hosaka and H. Toki, Phys. Lett. {\bf B322},  1  (1993).

\bibitem{bolo:schewe}
J. Schechter and H. Weigel, Resolving Ordering Ambiguities in the Collective
  Quantization by Particle Conjugation Constraints, [hep-ph/9410321], 1994.

\bibitem{bolo:kari}
S.Kahana and G.Ripka, Nucl. Phys. {\bf A429},  962  (1984).

\bibitem{bolo:dama2}
R. Dashen and A. Manohar, Phys. Lett. {\bf B315},  438  (1993).

\bibitem{bolo:ab4}
A.Blotz {\it et~al.}, Nucl. Phys. {\bf A555},  765  (1993).

\bibitem{bolo:ab6}
A.Blotz, M. Polyakov, and K. Goeke, Phys. Lett. {\bf B302},  151  (1993).


\end{thebibliography}


%
%
\begin{figure}[t]
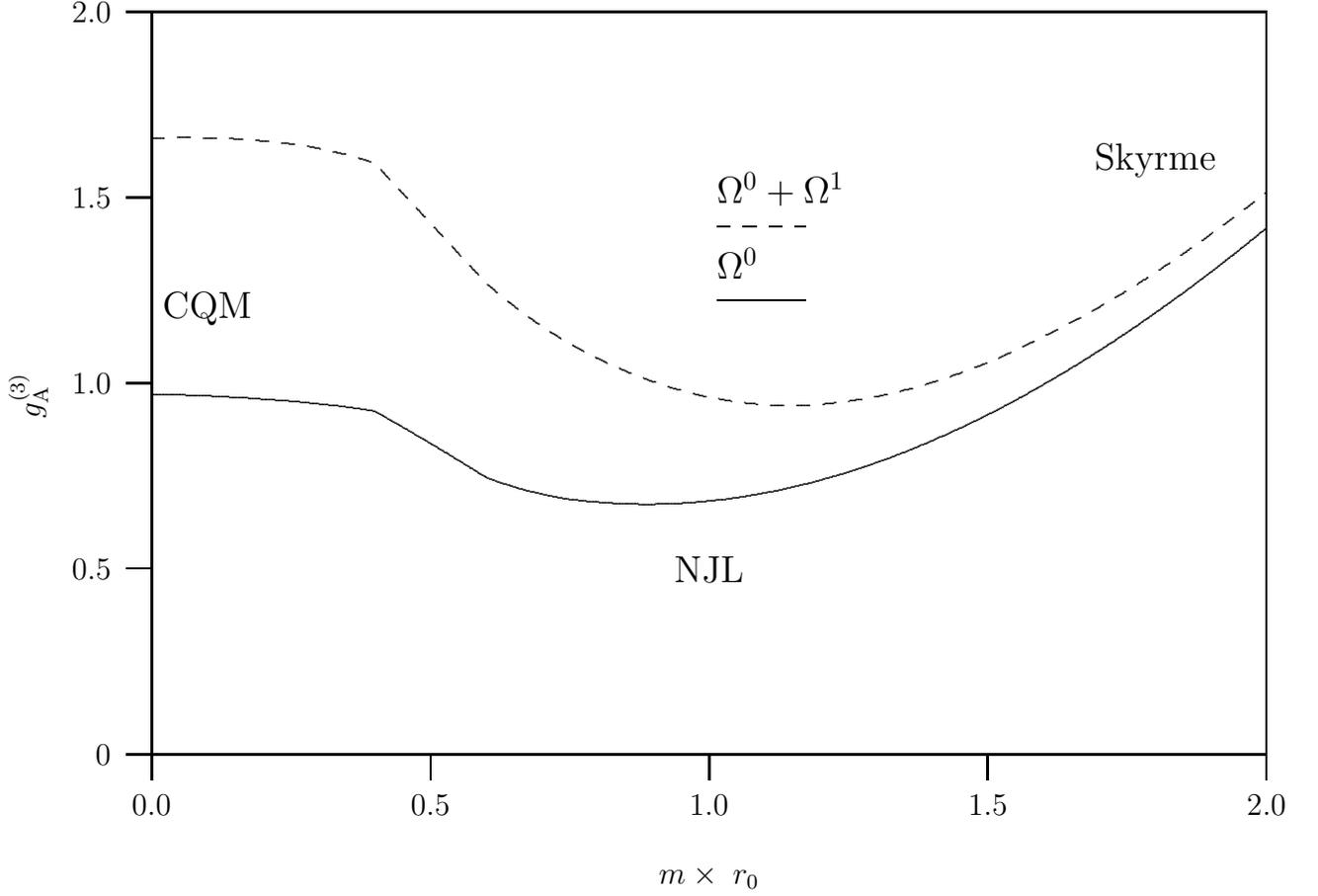

\def\rottext{\special{ps:currentpoint currentpoint translate -90 rotate
neg exch neg exch translate}}
\def\rotatetext#1{
\rottext\hbox
 to0pt{\hss#1\hss}
}
%
$$ \beginpicture
\setcoordinatesystem units <75mm,50mm>
\setplotarea x from 0  to 2, y from 0 to 2
\axis bottom label {$m\times \;r_0$} ticks
      numbered from 0 to 2 by  0.5
      unlabeled short quantity  5  /
\axis top /
\axis right /
\axis left ticks
          withvalues  0    0.5 1.0 1.5  2.0             /
          at          0    0.5 1.0 1.5  2.0             /
          unlabeled short quantity  5  /
\put{\rotatetext{ $g_{\rm A}^{(3)}$ }} at -0.2   1
\putrule from 0 0.0 to 2 0.0
\put{{\scriptsize
 \beginpicture \setcoordinatesystem units <1cm,1cm>
 \put{\large $\Omega^0+\Omega^1$ }[l] at 0  -1.5
 \setdashes
 \putrule from 0 -2 to 1.2 -2
 \put{\large $\Omega^0$  }[l] at 0  -2.5
 \setsolid
 \putrule from 0 -3  to 1.2 -3
 \endpicture
}}[bl] <3pt,3pt> at 1.0    1.2
%
\put{\large CQM} at 0.1   1.2
\put{\large NJL} at 1 0.5
\put{\large Skyrme} at 1.8  1.6
\inboundscheckon
\setquadratic
\setsolid
\plot
         0.0000        0.9694
         0.1000        0.9660
         0.2000        0.9565
         0.3000        0.9445
         0.4000        0.9250
         0.5000        0.8382
         0.6000        0.7464
         0.7000        0.7012
         0.8000        0.6796
         0.9000        0.6742
         1.0000        0.6826
         1.1000        0.7043
         1.2000        0.7388
         1.3000        0.7858
         1.4000        0.8447
         1.5000        0.9150
         1.6000        0.9961
         1.7000        1.0874
         1.8000        1.1884
         1.9000        1.2987
         2.0000        1.4176
/
\setdashes
\plot
        0.0000         1.6604
        0.1000         1.6611
        0.2000         1.6521
        0.3000         1.6323
        0.4000         1.5932
        0.5000         1.4317
        0.6000         1.2698
        0.7000         1.1537
        0.8000         1.0676
        0.9000         1.0042
        1.0000         0.9624
        1.1000         0.9425
        1.2000         0.9435
        1.3000         0.9638
        1.4000         1.0018
        1.5000         1.0556
        1.6000         1.1259
        1.7000         1.2047
        1.8000         1.2973
        1.9000         1.4006
        2.0000         1.5138
 /
\endpicture $$
\caption{The axial vector coupling constant $g_{\rm A}^{(3)}$ in dependence
of the size $r_0$ of the chiral field for the lowest
order $\Omega^0$ and next to leading order $\Omega^0+\Omega^1$
correction. The constituent quark mass is $m=370~\MeV$.
            }
\label{fig1}
\end{figure}
\end{document}